# Prediction of plastic anisotropy of textured polycrystalline sheets based on a new single crystal model

Nitin Chandola, Oana Cazacu, Benoit Revil-Baudard,

Department of Mechanical and Aerospace Engineering, University of Florida, REEF, 1350 N. Poquito Rd., Shalimar, FL 32579, USA.

## Abstract

In this paper, we predict the effect of texture on the anisotropy in plastic properties of polycrystalline metallic sheets. The constituent grain behavior is modelled using the new single crystal yield criterion developed by Cazacu, Revil, and Chandola (2017). For ideal texture components, the yield stress and plastic strain ratios can be obtained analytically. For the case of strongly textured sheets containing a spread about the ideal texture components, the polycrystalline response is obtained numerically on the basis of the same single-crystal criterion. It is shown that for textures with misorientation scatter width up to 25º, the numerical predictions are very close to those obtained analytically for an ideal texture. Furthermore, irrespective of the number of grains in the sample, Lankford coefficients have finite values for all loading orientations. Illustrative examples for sheets with textures containing a combination of few ideal texture components are also presented. The simulations of the predicted polycrystalline behavior based on the new description of the plastic behavior of the constituent grains capture the influence of individual texture components on the overall degree of anisotropy. The polycrystalline simulation results are also compared to analytical estimates obtained using the closed-form formulas for the ideal components present in the texture in conjunction with a simple law of mixtures. The analytical estimates show the same trends as the simulation results. Therefore, the trends in plastic anisotropy of the macroscopic properties can be adequately estimated analytically.

1. Introduction

Description of the plastic deformation of textured polycrystalline materials using advanced analytical orthotropic yield criteria that capture with accuracy the anisotropy in mechanical response of the metal in bulk have led to significant advances in metal technology. Examples of yield criteria for textured polycrystalline materials that are defined for three-dimensional loadings include Hill (Hill, 1948), Barlat (Barlat, 1987), Cazacu and Barlat (Cazacu and Barlat, 2001; Cazacu and Barlat, 2003; Cazacu and Barlat, 2004), Barlat and collaborators (Barlat et al., 2005).

In the framework of crystal plasticity, the most widely used approach for determining the macroscopic plastic behavior is based on the Schmid law for activation of slip in the constituent grains and Taylor's assumption of homogeneous deformation of all crystals (Taylor, 1938). There is an immense body of literature and publications on the Taylor model, also called Taylor-Bishop-Hill (TBH) model, (Taylor, 1938; Bishop and Hill, 1951a, 1951b). For a review of the TBH theory the reader is referred to the enlightening contribution of Van Houtte et al., 2004.

While increasingly complex homogenization schemes have been proposed (e.g. see Tome et al., 1991), use of such models for solving large-scale boundary value problems is still limited, mainly due to the prohibitive computational cost (e.g. see Eykens et al., 2015).

Recently, Cazacu, Revil, and Chandola (Cazacu et al. 2017) developed an analytical yield criterion for cubic single crystals. It is represented by a function which is $C^2$ differentiable for any three-dimensional stress states, and it accounts for the symmetries of the crystal. It involves four anisotropy coefficients and as such has added flexibility compared to the classical Schmid law or the regularized form of Schmid law (Arminjon, 1991). Specifically, the yield criterion (Cazacu et al. 2017) accounts for the differences in yield stress anisotropy between single-crystals (e.g. it captures the different relative ordering of the yield stresses as a function of the crystallographic direction of loading in single crystal copper as compared to aluminum single crystal).

It has been long recognized that the anisotropy in Lankford coefficients (plastic strain ratios or r-values) is related to the metal drawing performance and as such of interest to metallurgists engineers, and designers of metal forming (e.g. for aluminum single crystals see mechanical data and cup drawing test results reported by Carpenter and Elam, 1921 and Tucker, 1961 respectively; for polycrystalline aluminum sheets, see for example Lequeu et al., 1987, Barlat et al., 2004, Banabic et al., 2007, and others).

In this paper, using the new yield criterion (Cazacu et al. 2017) for describing the plastic behavior of the constituent crystals, we study the effect of texture on the plastic anisotropy of polycrystalline metallic sheets. Specifically, we predict the anisotropy in uniaxial yield stresses and Lankford coefficients (r-values) in metallic sheets containing the following texture components that are commonly observed experimentally: $\{100\}\langle 001\rangle$ (Cube), $\{110\}\langle 001\rangle$ (Goss), $\{112\}\langle 11\bar{1}\rangle$ (Copper), and $\{\bar{2}1\bar{1}\}<011>$. First, results are presented for sheets having one texture component specified in terms of a Gaussian distribution of misorientations with scatter width ranging from 0º (i.e. ideal texture) up to 45º around the respective ideal texture component (i.e. close to random texture). It is shown that for ideal textures, using the single-crystal criterion of Cazacu et al. (2017) the directional dependence of the yield stress and Lankford coefficients can be calculated analytically. For the case of strongly textured sheets containing a spread about the ideal texture components, the polycrystalline response is calculated numerically using the same single-crystal criterion of Cazacu et al.( 2017) for the description of the behavior of the constituent grains. The results of simulations of the polycrystalline behavior for textures with misorientation scatter width up to 25º about the ideal orientations are very close to the analytical ones, and it is predicted that Lankford coefficients have finite values for all loading orientations. Next, illustrative examples for sheets with textures containing a few ideal components are presented. These polycrystalline simulations results are compared to analytical estimates obtained using the closed-form formulas for the ideal components present in the texture in conjunction with a simple law of mixtures. We conclude with a summary of the main findings.

   2. **Constitutive Model**

Using the generalized invariants for cubic symmetry developed in Cazacu et al. (2017), one can construct yield criteria that are pressure-insensitive and satisfy the invariance requirements associated with the symmetries of each of the crystal classes of the cubic system. In this paper, we will use the single-crystal yield criterion developed for the hextetrahedral, gyroidal, and hexoctahedral cubic classes. This is motivated by the fact that most of the face centered cubic metals (e.g. copper; aluminum) belong to these crystal classes. Furthermore, for FCC crystals it can be assumed that the mechanical response in tension and compression is the same; therefore the following even function of the generalized invariants, proposed in Cazacu et al. (2017), will be further considered for the description of the plastic behavior of the constituent grains:

$$(J_2^C)^3 - c(J_3^C)^2 = k^6, \tag{1}$$

where $k$ denotes the yield limit in simple shear in any of the {100} crystallographic planes. In the above equation $c$ is a material constant that controls the relative importance of the generalized invariants of the stress deviator, $J_2^C$ and $J_3^C$, on yielding of the crystal.

In the coordinate system Oxyz associated with the <100> crystal axes, these generalized cubic invariants are expressed as:

$$\begin{aligned} J_2^C &= \frac{m_1}{6}\left[\left(\sigma_{xx}-\sigma_{yy}\right)^2 + \left(\sigma_{xx}-\sigma_{zz}\right)^2 + \left(\sigma_{zz}-\sigma_{yy}\right)^2\right] + m_2\left(\sigma_{xy}^2 + \sigma_{xz}^2 + \sigma_{yz}^2\right) \\ J_3^C &= \frac{n_1}{27}\left(2\sigma_{xx}-\sigma_{yy}-\sigma_{zz}\right)\left(2\sigma_{yy}-\sigma_{zz}-\sigma_{xx}\right)\left(2\sigma_{zz}-\sigma_{xx}-\sigma_{yy}\right) + 2n_4\sigma_{xy}\sigma_{xz}\sigma_{yz} \\ &\quad - \frac{n_3}{3}\left[\sigma_{yz}^2\left(2\sigma_{xx}-\sigma_{yy}-\sigma_{zz}\right) + \sigma_{xz}^2\left(2\sigma_{yy}-\sigma_{zz}-\sigma_{xx}\right) + \sigma_{xy}^2\left(2\sigma_{zz}-\sigma_{xx}-\sigma_{yy}\right)\right]. \end{aligned} \tag{2}$$

More details about the mathematical framework and the derivation of the expressions of these generalized invariants can be found in Cazacu et al. (2017).

Given that the criterion given by Eq. (1) with $J_2^C$ and $J_3^C$ given by Eq. (2), is a homogeneous function in stresses, the yielding response is the same if the coefficients $m_1, m_2, n_1, n_3, n_4$ are replaced by $\alpha m_1, \alpha m_2, \alpha n_1, \alpha n_3, \alpha n_4$, with $\alpha$ being an arbitrary positive constant. Therefore, without loss of generality one of the parameters, for example $m_1$, can be set equal to unity. Accordingly, the effective stress, $\bar{\sigma}$, associated to this single-crystal criterion is:

$$\bar{\sigma} = \frac{3}{\left(27-4cn_1^2\right)^{1/6}} \left\{ \begin{matrix} \left[\frac{1}{2}\left(\sigma_{xx}'^2+\sigma_{yy}'^2+\sigma_{zz}'^2\right) + m_2\left(\sigma_{xy}'^2+\sigma_{xz}'^2+\sigma_{yz}'^2\right)\right]^3 \\ -c\left[n_1\sigma_{xx}'\sigma_{yy}'\sigma_{zz}' - n_3\left(\sigma_{zz}'\sigma_{xy}'^2+\sigma_{xx}'\sigma_{yz}'^2+\sigma_{yy}'\sigma_{xz}'^2\right) + 2n_4\sigma_{xy}'\sigma_{xz}'\sigma_{yz}'\right]^2 \end{matrix} \right\}^{1/6}, \tag{3}$$

with $\sigma'$ denoting the Cauchy stress deviator.

Therefore, the single-crystal yield criterion involves only five independent parameters: $m_2, n_1, n_3, n_4$ and $c$. The coefficient $n_1$ has a clear physical significance being directly expressible in terms of the ratio between the yield limits in uniaxial tension along <100> directions and the yield limit in simple shear i.e. $k$. The remaining coefficients $m_2, n_3, n_4$ and $c$ can be determined from the tensile yield stresses along four different orientations. More details concerning the identification procedure can be found in Cazacu et al. (2017).

It is worth noting that the single-crystal criterion (1) is expressed by a differentiable function of class $C^2$ for any stress state. Assuming associated flow rule, the plastic strain-rate tensor, $\mathbf{d}^\mathbf{P}$, can be easily calculated as:

$$\mathbf{d}^\mathbf{P} = \dot{\lambda} \frac{\partial \bar{\sigma}}{\partial \boldsymbol{\sigma}} , \tag{4}$$

where $\dot{\lambda}$ is the plastic multiplier, and $\bar{\sigma}$ is given by Eq.(3).

The equivalent stress $\bar{\sigma}_{poly}$ of the polycrystalline material as a function of the applied stress tensor $\boldsymbol{\sigma}$, expressed in the loading frame, is:

$$\bar{\sigma}_{poly}(\boldsymbol{\sigma}) = \frac{1}{N} \sum_{i=1}^{N} \bar{\sigma}^i_{grain} \left( \mathbf{R}_i^T \boldsymbol{\sigma} \mathbf{R}_i \right) , \tag{5}$$

with N being the number of grains considered in the polycrystalline material, $\bar{\sigma}^i_{grain}$ is the effective stress of any given grain $i$, given by Eq. (3), and $\mathbf{R}_i$ is the transformation matrix for passage from the crystal axes of the grain $i$ to the loading frame.

Therefore, the plastic strain-rate deviator $\mathbf{D}^p$ of the polycrystalline material, expressed in the loading frame, is:

$$\mathbf{D}^\mathbf{p} = \frac{\dot{\lambda}}{N} \sum_{i=1}^{n} \mathbf{R}_i \frac{\partial \bar{\sigma}^i_{grain} \left( \mathbf{R}_i^T \boldsymbol{\sigma} \mathbf{R}_i \right)}{\partial \left( \mathbf{R}_i^T \boldsymbol{\sigma} \mathbf{R}_i \right)} \mathbf{R}_i^T \tag{6}$$

In the following, using the single crystal model for the description of the plastic behavior of the constituent grains (Eq. (3)-(6) )we predict the anisotropy of the plastic flow properties in uniaxial tension of polycrystalline materials containing texture components commonly observed experimentally. Specifically, we predict the effect of texture on the variation of the yield stresses $\sigma(\alpha)$ and strain-rate ratios $r(\alpha)$ with the orientation $\alpha$ of the loading axis.

We recall that by definition, the Lankford coefficient $r(\alpha)$ is the ratio between the in plane transverse strain-rate, $D_{22}^p$, and the through-thickness strain-rate, $D_{33}^p$, under uniaxial loading in a direction at angle $\alpha$ with respect to a reference direction in the plane of the polycrystalline sheet. In the Cartesian frame ($\mathbf{e}_1$, $\mathbf{e}_2$, $\mathbf{e}_3$) associated with the applied loading,

$$r(\alpha) = \frac{D_{22}^p}{D_{33}^p} . \qquad (7)$$

For all textures, the calculations are done assuming the same set of values for the parameters $m_2, n_1, n_3, n_4, c$, characterizing the plastic behavior of the constituent crystals (see Eq. (3)). These numerical values are: $m_2$=0.38, $n_1$=0.98, $n_3$=0.04, $n_4$=0.08, $c = 2.3$, and are representative of aluminum alloys.

### 3. Prediction of the yield stress and Lankford coefficients variation for selected ideal texture components

In polycrystalline metallic sheets, the crystals are not randomly oriented, but are distributed along preferred orientations that result from rotations that occur during processing. For a given fabrication process, the textures that develop contain one or several ideal components. The ideal texture components that will be considered in this paper are: {100}<001> (Cube), {110}<001> (Goss), {112}<111> (Copper), and $\{\bar{2}1\bar{1}\}<011>$. Experimentally a spread is generally observed around the various ideal texture components. Following Bunge, (2013), the following function is considered for the distribution of an orientation $g_0$:

$$f(g) = f(g_0)\exp(-\omega^2 / \omega_0^2). \tag{8}$$

In the above equation, $\omega$ is the rotation angle responsible for the spread about the ideal orientation $g_0$, and $\omega_0$ is the scatter width.

### 3.1 Cube-textured polycrystalline material

Let us first consider the case of a polycrystalline material containing only the {100}<001> (Cube texture) component. For an ideal texture ($\omega_0 = 0$°), the <100> axes of all the constituent crystals are aligned with the texture axes. Under in-plane uniaxial tension, the only non-zero components of the stress, referred to the Oxyz crystal axes, are: $\sigma_{xx} = \sigma(\alpha)\cos^2\alpha$, $\sigma_{yy} = \sigma(\alpha)\sin^2\alpha$, $\sigma_{xy} = \sigma(\alpha)\sin\alpha\cos\alpha$, with $\sigma(\alpha)$ being the yield stress along the in-plane direction $\alpha$. The variation of $\sigma(\alpha)$ with the tensile loading orientation can be obtained analytically by substituting the above stresses in the expression (3) of the effective stress associated to the single-crystal yield criterion of Cazacu et al.(Cazacu et al. 2017) as:

$$\frac{\sigma(\alpha)}{\sigma(0)} = \frac{\left(27 - 4cn_1^2\right)^{1/6}}{\left\{27\left[1 + 3(m_2 - 1)\sin^2\alpha\cos^2\alpha\right]^3 - c\left[2n_1 + 9(n_3 - n_1)\sin^2\alpha\cos^2\alpha\right]^2\right\}^{1/6}}. \tag{9}$$

Likewise, use of Eq. (4) leads to the following expression for the variation of the plastic strain-rate ratios $r(\alpha)$ with the loading orientation, $\alpha$,

$$r(\alpha) = -\frac{\sin^2\alpha\,\dfrac{\partial\bar{\sigma}}{\partial\sigma_{xx}} - \sin 2\alpha\,\dfrac{\partial\bar{\sigma}}{\partial\sigma_{xy}} + \cos^2\alpha\,\dfrac{\partial\bar{\sigma}}{\partial\sigma_{yy}}}{\dfrac{\partial\bar{\sigma}}{\partial\sigma_{xx}} + \dfrac{\partial\bar{\sigma}}{\partial\sigma_{yy}}}, \tag{10}$$

with $\bar{\sigma}$ being the effective stress given by Eq. (3). Irrespective of the values of the parameters $m_2, n_1, n_3, n_4, c$, note that the material symmetries are correctly captured. Indeed, Eq. (9)-(10) predict that the response is identical under rotations of 90º about the normal direction, and in particular: $\sigma(0º) = \sigma(90º)$ and $r(0º) = r(90º) = 1$.

For polycrystalline sheets with textures of increased scatter width ranging from $\omega_0 = 10º$ up to 45º from the ideal {100}<001> cube texture, simulations were done using the polycrystal model (Eq. (3)-(6)) for samples of 400 crystals. As an example, in Fig. 1 are shown the {111} pole figures for textures with a width spread about the ideal texture $\omega_0 = 25º$, 35º, and 45º, respectively (see Eq. (8)).

The evolution of yield stress ratios $\sigma(\alpha)/\sigma(0)$ and Lankford coefficients $r(\alpha)$ with the loading direction $\alpha$ for these cube-textured sheets according to the polycrystal model with the given values for the parameters describing the constituent grain behavior ($m_2=0.38$, $n_1=0.98$, $n_3=0.04$, $n_4=0.08$, $c = 2.3$) is shown in Fig.2 (pole figures for the textures are given in Fig. 1). On Fig. 2 are also plotted the simulation results corresponding to a texture with $\omega_0 = 30º$ and the analytical estimates for the ideal cube texture ($\omega_0 = 0º$) calculated using Eq.(9)-(10).

Note that irrespective of the scatter width $\omega_0$ about the ideal texture both the predicted macroscopic yield stresses and r-values vary smoothly with the loading orientation (see Fig.2). The curves are practically identical for $\omega_0$ ranging from $\omega_0 = 0º$ (ideal texture) to $\omega_0 = 25º$, and very close to the ones corresponding to $\omega_0 = 30º$, as it should be given the very small variation between textures (see also Fig. 1(a)). Note also that for a very small spread from the ideal cube texture, the predicted yield stresses and r-values are almost identical under rotations of 90º about the sheet normal, meaning that the model correctly predicts the strong symmetries associated with the given textures (see also Eq.(9)). For the given values of the parameters characterizing the grains behavior, minima in yield stresses are along the 0º and 90º orientations and there is only one peak which corresponds to uniaxial loading at 45º. The predicted variation of the Lankford coefficients with the orientation is such that there is only one minimum corresponding to uniaxial tension along the 45º in-plane direction.

For a texture with $\omega_0 = 35°$, the directional dependence of the macroscopic plastic properties is similar to that predicted analytically for an ideal texture, but the anisotropy is less pronounced. Also, the minimum r-value is slightly shifted, corresponding to a tensile loading orientation of less than 45°. On the other hand, if the material has a texture with a very large scatter spread around the ideal cube texture ($\omega_0 = 45°$), which is close to a random texture ( see Fig. 1(c)), the new polycrystal model predicts that irrespective of the loading orientation both $\sigma(\alpha)/\sigma(0)$ and the r-values are close to unity (i.e. close to isotropic response).

It is well documented that for an ideal ($\omega_0 = 0°$) {100}<001> texture the yield stress variation with the loading orientation according to the TBH model displays two cusps while the Lankford coefficients are not defined for the 0° and 90° tensile loadings (e.g. see Lequeu et al. 1987) Only when the texture is characterized by a larger spread, the predicted variation in both the macroscopic yield stresses and Lankford coefficients is smooth (see Fig. 3, after Lequeu et al. 1987).

### 3.2 Goss texture $\{110\}\langle 001 \rangle$

For an ideal $\{110\}\langle 001 \rangle$ textured sheet (Goss texture), using the single-crystal yield criterion of Cazacu et al. (2017), an analytic formula for the evolution of the uniaxial yield stress ratio $\sigma(\alpha)/\sigma(0)$ with the loading direction $\alpha$ in the plane of the sheet can be obtained by referring the applied stress tensor to the crystal axes and further substituting the respective components in Eq.(3). The variation of the tensile yield stresses as a function of the loading direction $\alpha$ and the coefficients $m_2, n_1, n_3, n_4, c$ is:

$$\frac{\sigma(\alpha)}{\sigma(0)} = \frac{2\left(27-4cn_1^2\right)^{1/6}}{\left\{\left[12+9(m_2-1)\left(1+3\cos^2\alpha\right)\sin^2\alpha\right]^3 - c\left[16n_1+54\left(n_1-3n_3+2n_4\right)\sin^4\alpha\cos^2\alpha+18\left(n_3-n_1\right)\sin^2\alpha\left(1+3\cos^2\alpha\right)\right]^2\right\}^{1/6}} \quad (11)$$

Likewise, the variation of the plastic strain-rate ratios $r(\alpha)$ with the orientation $\alpha$ can also be obtained analytically as a function of the coefficients $m_2, n_1, n_3, n_4, c$ by making use of Eq. (4):

$$r(\alpha) = \frac{(3\cos^2\alpha - 2)\left(\dfrac{\partial \bar{\sigma}}{\partial \sigma_{xx}} + \dfrac{\partial \bar{\sigma}}{\partial \sigma_{yy}}\right) - 2\cos^2\alpha \dfrac{\partial \bar{\sigma}}{\partial \sigma_{xy}} - \sqrt{2}\sin(2\alpha)\dfrac{\partial \bar{\sigma}}{\partial \sigma_{xz}}}{\dfrac{\partial \bar{\sigma}}{\partial \sigma_{xx}} + \dfrac{\partial \bar{\sigma}}{\partial \sigma_{yy}} + 2\dfrac{\partial \bar{\sigma}}{\partial \sigma_{xy}}}, \qquad (12)$$

where $\bar{\sigma}$ is the plastic potential given by Eq.(3) and its derivatives are expressed in terms of the stress components in the crystal axes. Note that irrespective of the values of the parameters $m_2, n_1, n_3, n_4, c$, the model correctly accounts for the material symmetries, namely that $r(0°) = 1$.

In Fig.5 is shown the predicted variations of macroscopic plastic properties with the loading orientation according to the analytical formulas (11)-(12) for the given textures (the (111) pole figures are shown in Fig.4). The effect of texture, namely of the spread $\omega_0$ with respect to the ideal Goss texture, on the directional dependence of the same macroscopic properties (yield stresses, plastic strain ratios) obtained using the polycrystalline model are plotted on the same figure for comparison. For the texture with $\omega_0 = 30°$, the $\sigma(\alpha)/\sigma(0)$ vs. $\alpha$ and $r(\alpha)$ vs. $\alpha$ variations are only slightly different than the respective curves for the ideal Goss texture.

For the given set of numerical values of the coefficients $m_2, n_1, n_3, n_4, c$, (see Section 1) the absolute maximum in yield stresses shifts from the loading orientation at ~50° with respect to the reference direction to ~40° while in the variation of $r(\alpha)$ the peak shifts to $\alpha$ ~80° orientation, the anisotropy remaining as strong (e.g. r(80°) still about eight times larger than r(0°)). Note that for a texture with $\omega_0 = 35°$, the overall trends in the directional dependence of the plastic properties are similar to those corresponding to a texture with $\omega_0 = 30°$, but the anisotropy in r-values is much less pronounced, r(90°) being five times larger than r(0°)=1. Obviously, for the texture with $\omega_0 = 45°$, the predicted response is very close to the isotropic one (see also Fig. 5).

### 3.3 Copper texture $\{112\}\langle 11\bar{1}\rangle$

For an ideal $\{112\}<11\bar{1}>$ textured sheet (Copper texture), an analytic formula for the variation of the uniaxial yield stress ratio $\sigma(\alpha)/\sigma(0)$ with the loading direction $\alpha$ in the plane of the sheet can be obtained by referring the applied stress tensor to the crystal axes and further substituting the respective components in Eq.(3):

$$\frac{\sigma(\alpha)}{\sigma(0)} = \frac{2\left(27 - 4cn_1^2\right)^{1/6}}{\left\{\begin{array}{l}\left[12 + 3(m_2 - 1)\left(4\cos^4\alpha + 3\sin^4\alpha\right)\right]^3 - \\ c\left[16n_1 + 2(n_1 - 3n_3 + 2n_4)\sin^2\alpha\left(5\sin^2\alpha - 2\right) + 6(n_3 - n_1)\left(4\cos^4\alpha + 3\sin^4\alpha\right)\right]^2\end{array}\right\}^{1/6}}$$
(13)

Similarly, the dependence of the plastic strain-rate ratios $r(\alpha)$ with the loading orientation $\alpha$ can also be obtained analytically as a function of the coefficients $m_2, n_1, n_3, n_4, c$ characterizing the constituent crystals plastic behavior. For the given set of numerical values of these parameters the predicted variation is shown in Fig.7. For textures with a scatter spread $\omega_0$ from the ideal copper texture up to 30º, the response is almost identical to that obtained for the ideal texture ($\omega_0$ =0º) by applying the analytical formula. For the texture with $\omega_0$ =35º (see Fig.6 for the {111} pole figures), the trends in the directional dependence of the macroscopic plastic properties is similar, the maximum Lankford coefficient being obtained in uniaxial tension at an orientation $\alpha$ about 20º (as compared to the ideal texture for which the predicted maximum corresponds to $\alpha \sim$ 30º). For the texture with $\omega_0$ =45º (see Fig. 6), the predicted response is close to the isotropic one (predicted minimum in r-values is 0.843 while the maximum is 1.014).

### 3.4 $\{\bar{2}1\bar{1}\}<011>$ texture component

For an ideal texture $\{\bar{2}1\bar{1}\}<011>$, using the same criterion given by Eq. (1) for the description of the plastic deformation of each grain, under uniaxial tension along an axis at orientation $\alpha$, the yield stress is:

$$\frac{\sigma(\alpha)}{\sigma(0)} = \frac{\left[(9m_2+3)^3 - 256cn_1^2\right]^{1/6}}{\left\{\begin{array}{l}\left[12+3(m_2-1)(3\cos^4\alpha+4\sin^4\alpha)\right]^3 - \\ c\left[16n_1+2(n_1-3n_3+2n_4)(5\cos^2\alpha-2)\sin^2\alpha+6(n_3-n_1)(3\cos^4\alpha+4\sin^4\alpha)\right]^2\end{array}\right\}^{1/6}}$$

(14)

The variation of the Lankford coefficients with the loading orientation $\alpha$ is given by:

$$r(\alpha) = \begin{bmatrix} -(3\sin^2\alpha+\sqrt{6}\sin 2\alpha)\frac{\partial\bar{\sigma}}{\partial\sigma_{xx}} - 2\sqrt{6}\sin 2\alpha\frac{\partial\bar{\sigma}}{\partial\sigma_{yy}} \\ +(1-5\cos 2\alpha)\frac{\partial\bar{\sigma}}{\partial\sigma_{yz}} \\ +4\cos^2\alpha\left(\frac{\partial\bar{\sigma}}{\partial\sigma_{xy}} - \frac{\partial\bar{\sigma}}{\partial\sigma_{xz}}\right) \\ -\sqrt{6}\sin 2\alpha\left(\frac{\partial\bar{\sigma}}{\partial\sigma_{xy}} + \frac{\partial\bar{\sigma}}{\partial\sigma_{xz}}\right) \end{bmatrix} \left[3\left(\frac{\partial\bar{\sigma}}{\partial\sigma_{xx}}\right) + 2\left(\frac{\partial\bar{\sigma}}{\partial\sigma_{xz}} - \frac{\partial\bar{\sigma}}{\partial\sigma_{xy}} - \frac{\partial\bar{\sigma}}{\partial\sigma_{yz}}\right)\right]^{-1}$$

(15)

In the above expression, $\bar{\sigma}$ is given by Eq.(3), and its derivatives are expressed in terms of the stress components in the crystal axes.

In Fig. 9 are shown the predicted macroscopic mechanical properties in uniaxial tension for the material with ideal texture calculated using the analytical formulas (Eq.(14)-(15)) and the polycrystalline simulation results for the textures corresponding to Gaussian distributions of scatter width $\omega_0$=30°, 35°, and 45° with respect to the ideal $\{\bar{2}1\bar{1}\}<011>$ texture (see Fig. 8 for the $\{111\}$ pole figures of the selected textures).

First, let us note that irrespective of the scatter spread $\omega_0$, ranging from ideal to 30º, it is predicted a moderate anisotropy in yield stresses, the variation of the yield stresses with the loading orientation being almost the same, with a minimum at the 45º loading orientation, and a maximum at $\alpha=90º$. For the texture with $\omega_0=35º$ about the ideal texture, the predicted trends are similar (same curvature) with the minimum yield stress at $\alpha \sim 20º$. As concerns the predicted anisotropy in Lankford coefficients (see Fig. 9(b)), the trends are the same for textures with $\omega_0$ ranging from ideal ($\omega_0 = 0º$) to $\omega_0=35º$. As the width spread $\omega_0$ increases the anisotropy is less pronounced (the maximum shifts towards $\alpha=51º$ loading orientation and the maximum r-value decreases). For a texture with $\omega_0=45º$, the predicted response is close to the isotropic one as it should be given the degree of randomization of this texture (see Fig. 8).

In summary, irrespective of the texture component considered the analytical formulas provide a very good estimate of the anisotropy in macroscopic plastic properties. Next, using the new yield criterion (Cazacu et al. 2017) (see Eq. (1)) for the description of the plastic behavior of the constituent grains, we will investigate the predicted mechanical response of strongly textured polycrystalline materials containing various combinations of ideal texture components.

## 4 Predictions of anisotropy of yield stresses and Lankford coefficients for textured sheets containing several components

Let us first consider a polycrystalline sheet with components spread around the {100}<001> (80% volume fraction) and {110}<001> (20% volume fraction) ideal orientation, respectively. The texture of the polycrystalline sheet is shown in Fig. 10. The results of numerical simulations using the proposed polycrystalline model (Eq. (5)-(6)) are compared with the macroscopic yield stress and plastic strain ratios obtained by using the analytical formulas for {100}<001> and {110}<001> ideal textures in conjunction with a simple law of mixtures (Fig.11).

Note that the analytical estimates are very close to the numerical polycrystalline simulations results obtained using the same criterion (i.e. Cazacu et al. 2017) for the description of the plastic behavior of the constituent grains. The shapes of both the $r(\alpha)$ and $\sigma(\alpha)/\sigma(0)$ curves are similar to those corresponding to the cube texture (compare Fig. 11 with Fig. 2). However, the presence of the Goss component (20% volume fraction) results in r(90°) larger than r(0°) and the minimum r-value is slightly larger than in the case of the ideal cube texture. Specifically, for the given values of the coefficients characterizing the behavior of the constituents grains, $r(90°) = 1.27 > r(0°)=1$ and the minimum r-value corresponds to tensile loading along a direction $\alpha = 39°$. Note also that $r(39°)=0.21$, which is double the minimum r-value for an ideal cube texture (see Fig. 2(b) minimum is $r(45°)=0.1$).

For a polycrystalline sheet with texture containing Copper and Goss components in the same proportion (see pole figures in Fig.12), the results of the polycrystalline numerical simulations and analytical estimates obtained using the formulas for each ideal texture component in conjunction with a law of mixtures are shown in Fig. 13. Note that the analytical predictions are very close to the numerical predictions.

It is interesting to note that the predicted $\sigma(\alpha)/\sigma(0)$ curve is almost flat indicating little variation in yield stresses for loading orientation $\alpha$ between 0° and 20°, the anisotropy becoming slightly more pronounced for loading directions between 20° and 70°, with a peak in yield stress around 50°, and very little difference between yield stresses for loadings between 70° and 90°. While the shape of the $\sigma(\alpha)/\sigma(0)$ curve is concave down thus closer to that of the ideal Goss component (see also Fig. 5(a)), the anisotropy is much less pronounced and similar to that of the ideal copper component (compare with Fig. 7). It is interesting to note that although $r(0°)=1$ as it is the case for an ideal Goss component, and in the $r(\alpha)$ vs. $\alpha$ curve there is an inflexion point between 40° and 50°, also observed in the r-value variation for an ideal Goss component (see Fig. 5(b)). However, the r-value predicted for tensile loading at $\alpha = 90°$ is much lower. The presence of the copper component in the texture, lowers the $r(90°)$ value from about r(90°)~8, in the case of an ideal Goss component (see Fig. 5(b)), to $r(90°) = 1.57$.

The last example presented is that of a polycrystalline sheet with texture shown in Fig. 14. While the dominant texture component (70% volume fraction) has a spread about {110}<001> Goss, the texture also contains a component with spread about $\{\bar{2}1\bar{1}\}<011>$ (30% volume fraction). Fig. 15 presents the predicted evolution of the macroscopic yield stresses and plastic strain ratios for this material obtained on the basis of the same yield criterion for the constituent grains i.e. the analytical estimate (based on the analytical formulas for each ideal component present in the texture) and the results of polycrystalline simulations. The analytical $\sigma(\alpha)/\sigma(0)$ vs. $\alpha$ variation is very close to the numerical one. The analytical $r(\alpha)$ vs. $\alpha$ and the polycrystalline simulation results are similar with $r(0º)=1$, very little variation in r-values for loading orientations $\alpha$ between 0º and 40º, inflection point at $\alpha=40º$ and a sharp upward trend as in the case of the ideal Goss component. It is to be noted that although analytically a higher r-value is predicted in the transverse direction ($\alpha=90º$) than the numerically predicted value, the analytical estimate captures the influence of the $\{\bar{2}1\bar{1}\}<011>$ component. Namely, it predicts that the $\{\bar{2}1\bar{1}\}<011>$ contributes to a significant decrease in anisotropy of the material. For example, the predicted *r*-value in the transverse direction is significantly lower than that corresponding to a spread about the ideal Goss component (compare Fig. 15(b) with Fig. 5(b)).

## 5 Conclusions

Using a new constitutive model developed by Cazacu, Revil, and Chandola (2017) for the description of the individual constituent grains, the effect of texture on macroscopic plastic properties has been investigated. This single-crystal yield criterion is expressed in terms of generalized stress invariants and as such automatically satisfies the intrinsic crystal symmetries. It is represented by a function $C^2$ differentiable for any three-dimensional loadings. Therefore, for ideal texture components, it is possible to derive analytic formulas for the variation of the macroscopic yield stress and Lankford coefficients with the in-plane loading direction.

For the case of strongly textured materials with a distribution of grain orientations with various spreads about the ideal texture components, the anisotropy of the polycrystalline response is simulated numerically on the basis of the same single-crystal criterion. An added advantage is that irrespective of the number of grains in the sample, Lankford coefficients have finite values for all loading orientations even for ideal texture components, i.e. there is no need to add random texture components to gauge the plastic properties of the polycrystal. Moreover, for textured materials with grain orientations spread around one ideal texture components $\omega_0 \leq 30º$, the simulation results of polycrystalline response obtained with samples of 400 grains are very close to those obtained analytically. Irrespective of the texture component considered, the analytical formulas provide a good estimate of the degree of in-plane anisotropy and its trends at a very low calculation cost.

When more than one ideal texture component exists in the material, polycrystalline simulations based on the new description of the plastic behavior of the constituent grains capture the influence of individual texture components on the overall plastic anisotropy of the polycrystal. Additionally, it was shown that the use of the analytical formulas for each ideal component in conjunction with a law of mixtures provides an adequate estimate of the in-plane anisotropy. The analytical estimates show the same trends as the simulation results in terms of both yield stress ratios and r-values. Therefore, the trends in plastic anisotropy of the macroscopic properties, and most importantly how the predominant texture components affect the deformation can be adequately estimated analytically using the described approach.

Figure(s)

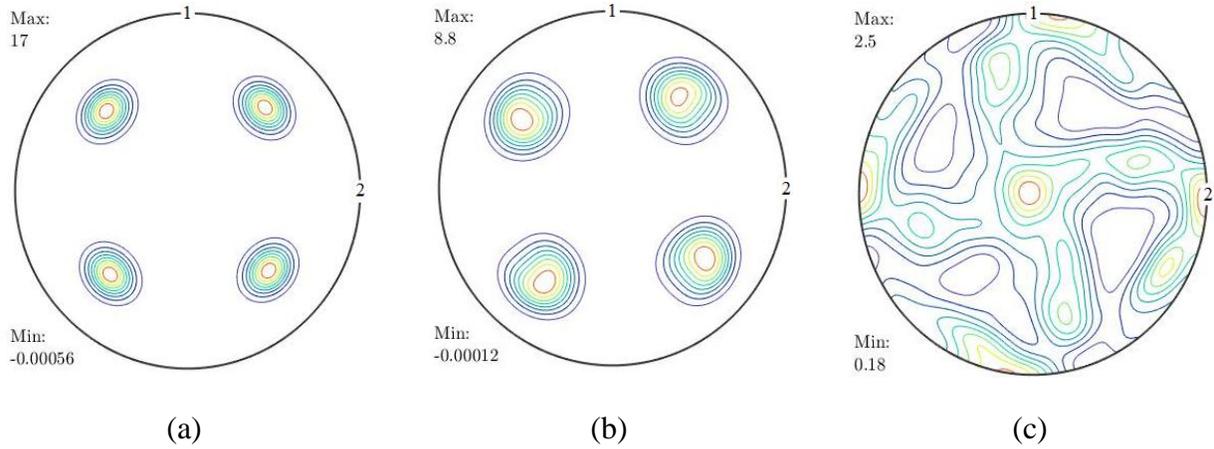

Fig. 1. (111) pole figures for textures corresponding to a series of Gaussian distributions of increasing scatter width, $\omega_0$, about the ideal {100}<001> texture: (a) $\omega_0=25^0$, (b) $\omega_0=35^0$, (c) $\omega_0=45^0$.

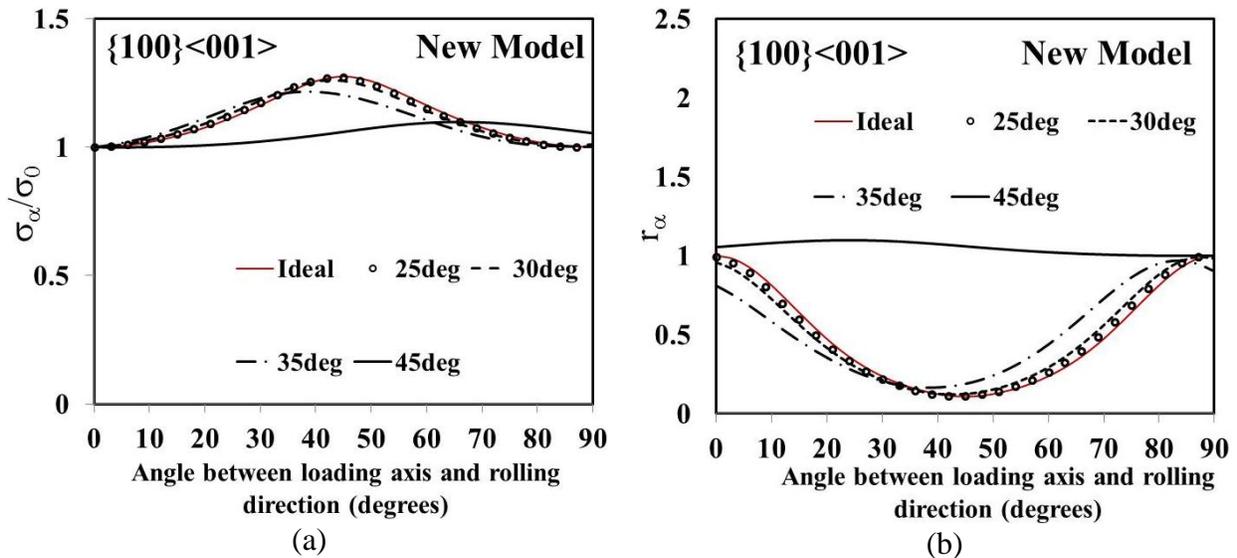

Fig. 2. Effect of initial texture on the anisotropy in (a) yield stress ratio $\sigma(\alpha)$ and (b) strain-ratio $r(\alpha)$ in the plane of the {100}<001> textured polycrystalline sheet predicted by the new polycrystal model. The textures for different scatter width $\omega_0$ are shown in Figure 1.

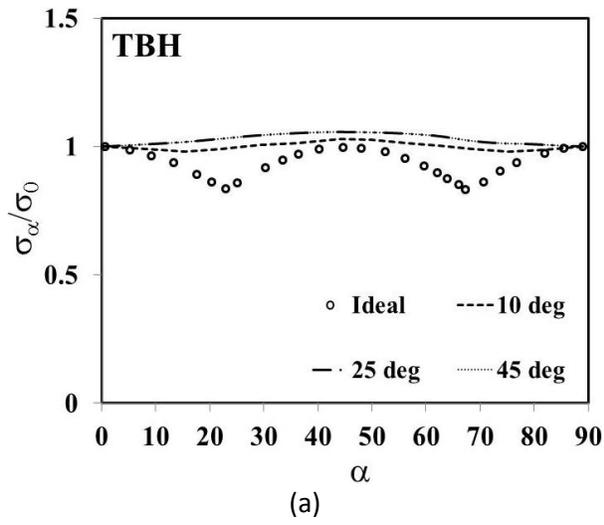 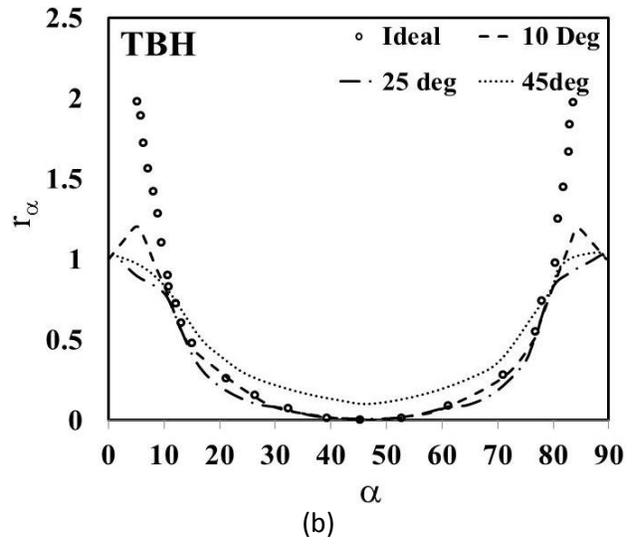

Fig. 3. Effect of texture on the anisotropy in: (a) yield stress ratio and (b) strain-ratio in the plane of the polycrystalline cube-textured sheet using the Taylor-Bishop-Hill approach (after Lequeu et al., 1987).

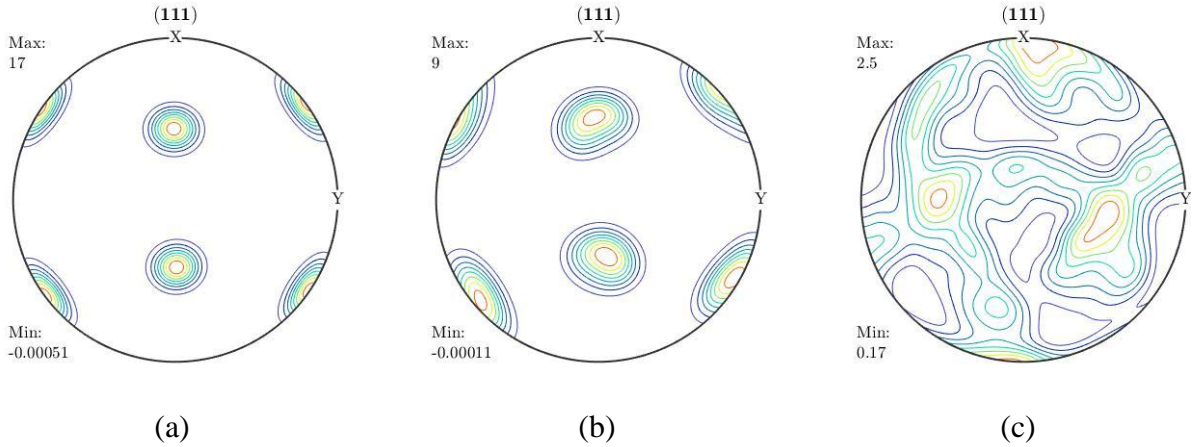

Fig. 4. (111) pole figures for textures corresponding to a series of Gaussian distributions of increasing scatter width $\omega_0$ about the ideal {110}<001> texture : (a) $\omega_0=25^0$, (b) $\omega_0=35^0$, (c) $\omega_0=45^0$.

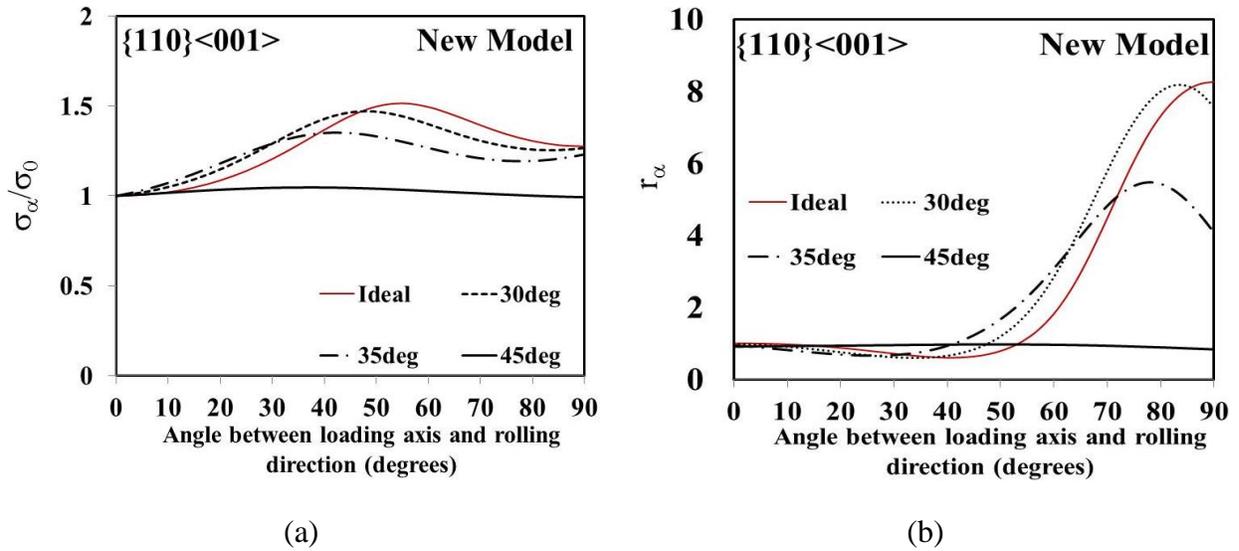

Fig. 5. Effect of texture on the anisotropy in (a) yield stress ratio and (b) strain-ratio in the plane of the polycrystalline sheet predicted by the new polycrystal model for {110}<001> texture. The textures for different scatter width $\omega_0$ are shown in Figure 4.

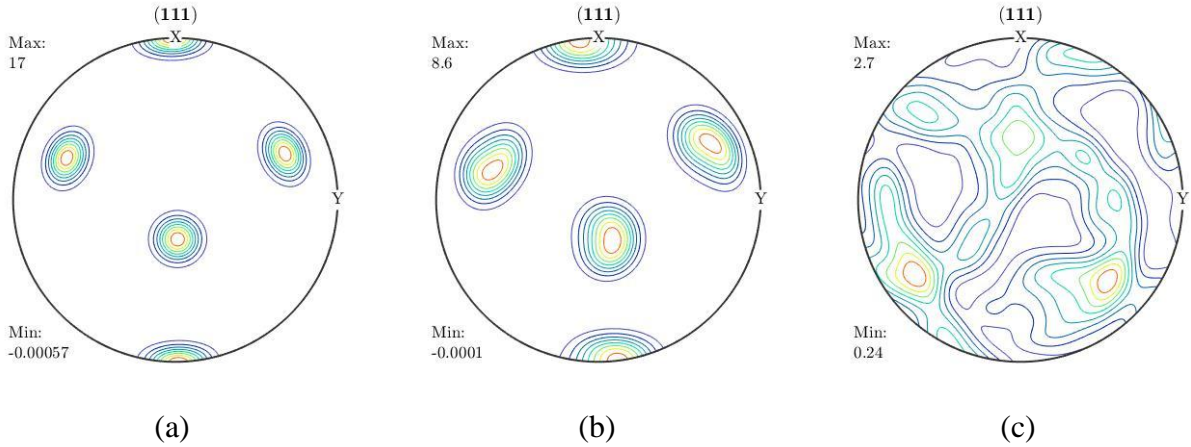

Fig. 6. (111) pole figures for textures corresponding to a series of Gaussian distributions of increasing scatter width $\omega_0$ about the ideal $\{112\}<1\bar{1}1>$ texture: (a) $\omega_0=25^0$, (b) $\omega_0=35^0$, (c) $\omega_0=45^0$.

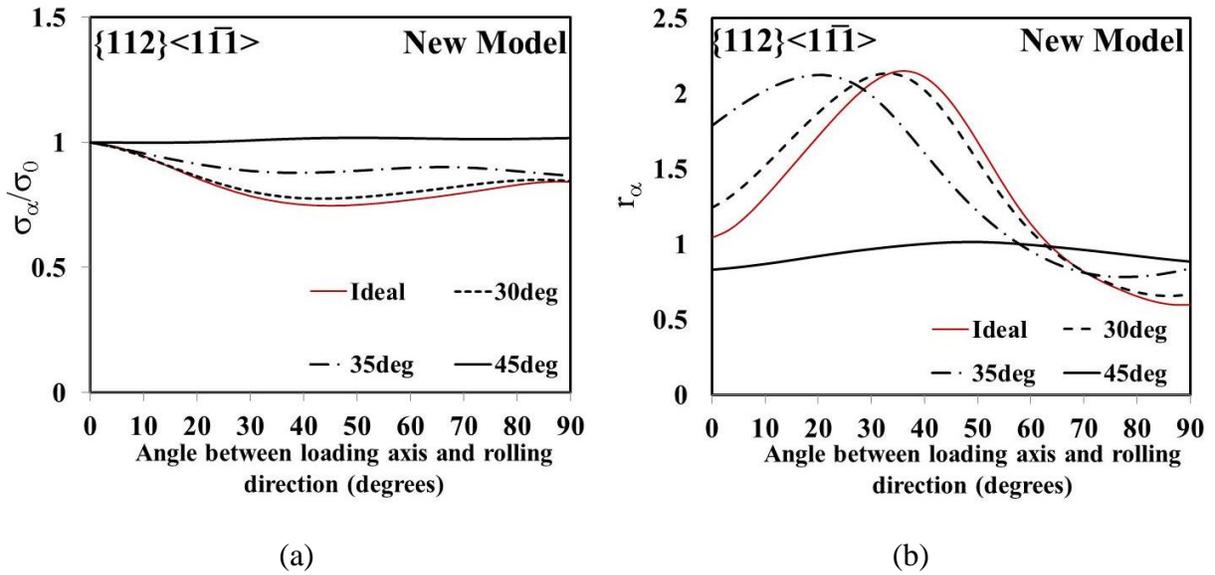

Fig. 7. Effect of texture on the anisotropy in (a) yield stress ratio and (b) strain-ratio in the plane of the polycrystalline sheet predicted by the new polycrystal model for $\{112\}<1\bar{1}1>$ texture. The textures for different scatter width $\omega_0$ are shown in Figure 6.

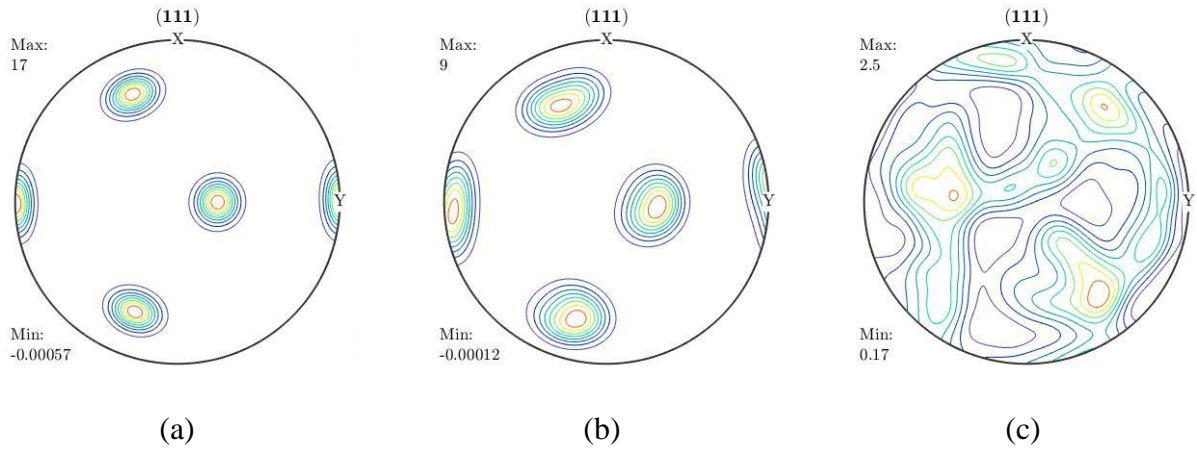

Fig. 8. (111) pole figures for textures corresponding to a series of Gaussian distributions of increasing scatter width $\omega_0$ about the ideal $\{\bar{2}\bar{1}1\}\langle011\rangle$ texture: (a) $\omega_0=25^0$, (b) $\omega_0=35^0$, (c) $\omega_0=45^0$.

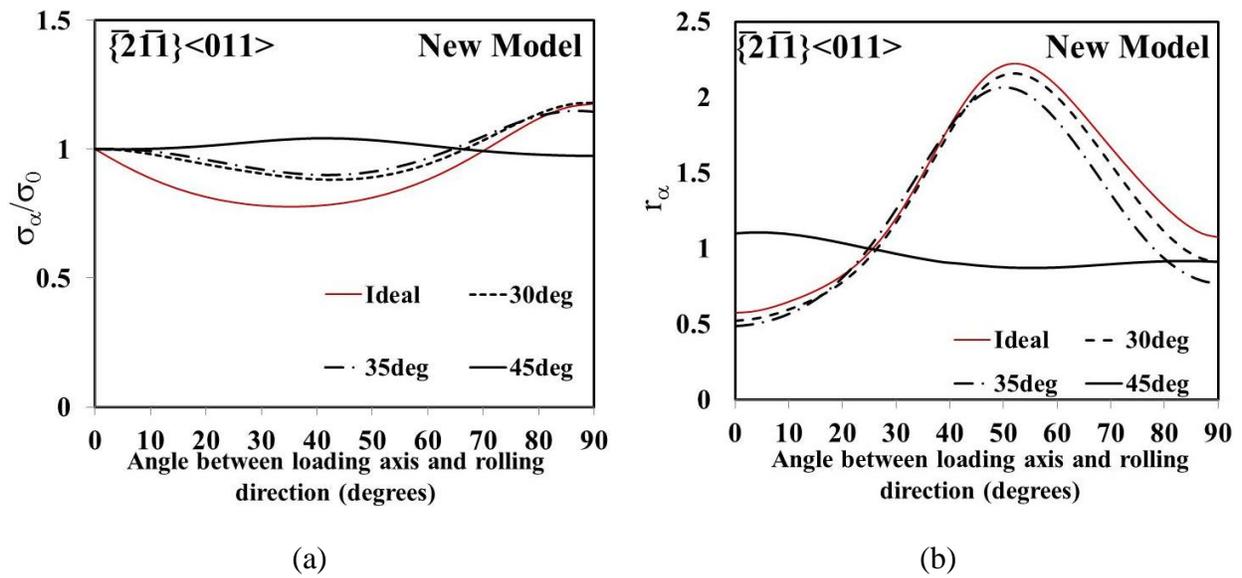

Fig. 9. Effect of texture on the anisotropy in (a) yield stress ratio and (b) strain-ratio in the plane of the polycrystalline sheet predicted by the new polycrystal model for $\{\bar{2}\bar{1}1\}\langle011\rangle$ texture. The textures for different scatter width $\omega_0$ are shown in Figure 8.

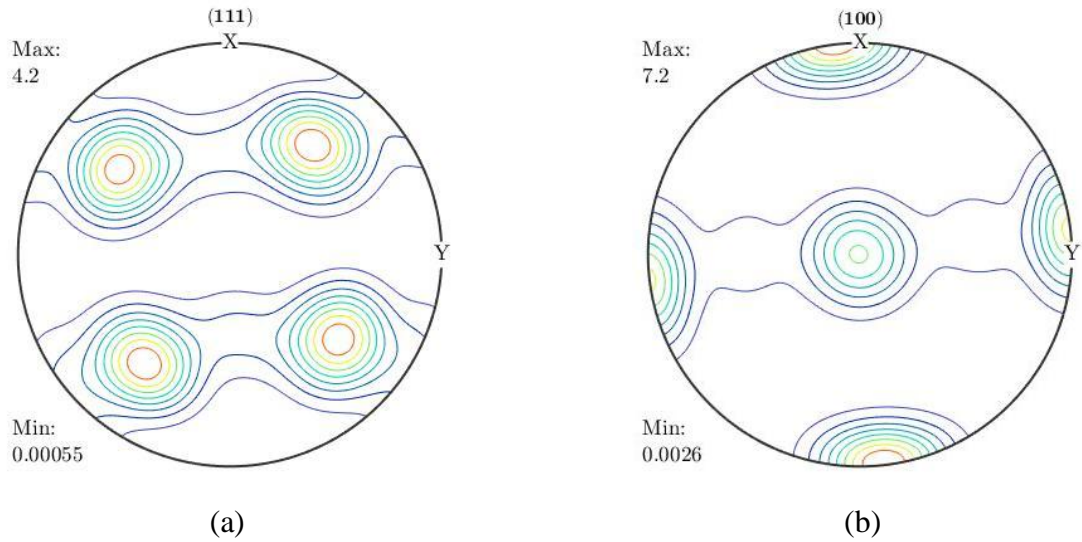

Fig.10. Pole figures for a polycrystal with mixture of 100}<001> component (80% volume fraction) and {110}<001> component (20% volume fraction): (a) (111) (b) (100).

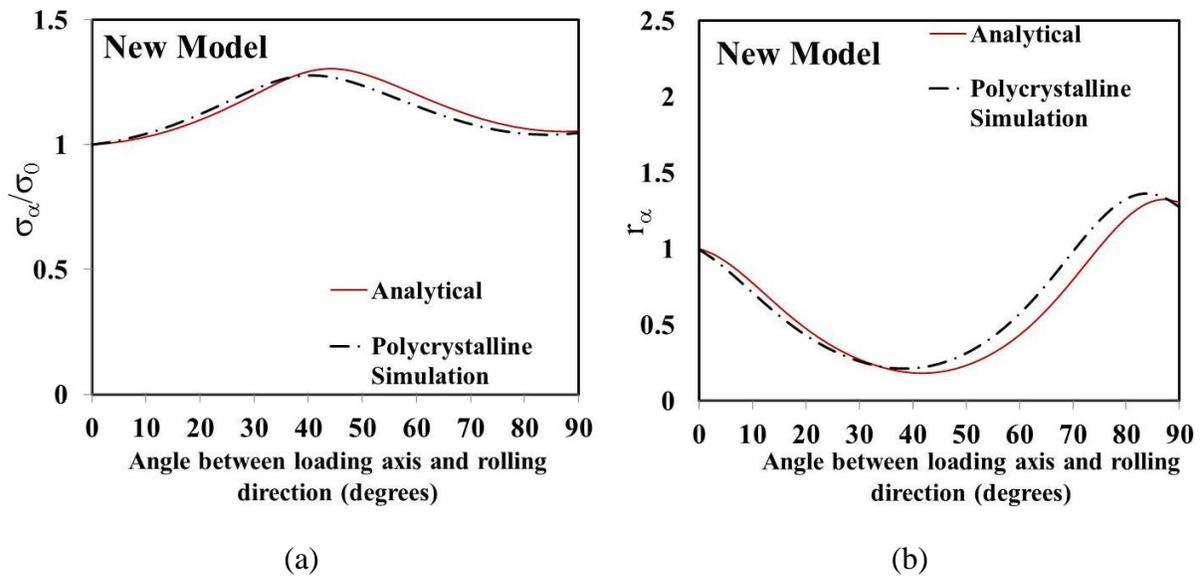

Fig. 11. Prediction of (a) yield stress ratio and (b) strain-ratio in the plane of the polycrystalline sheet predicted by the new polycrystal model for a strongly textured polycrystal with components spread around the {100}<001>(80%) and {110}<001>(20%) orientation. The texture is shown in Figure 10.

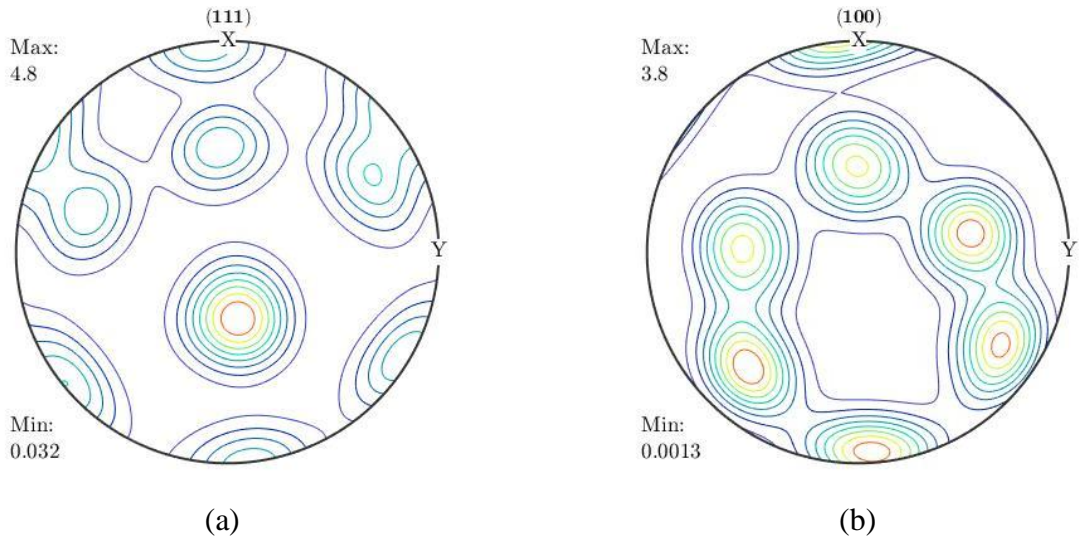

Fig.12. Pole figures for a polycrystal with mixture of {112}<1$\bar{1}$1> components (50%) and {110}<001> component (50%) (a) (111) (b) (100).

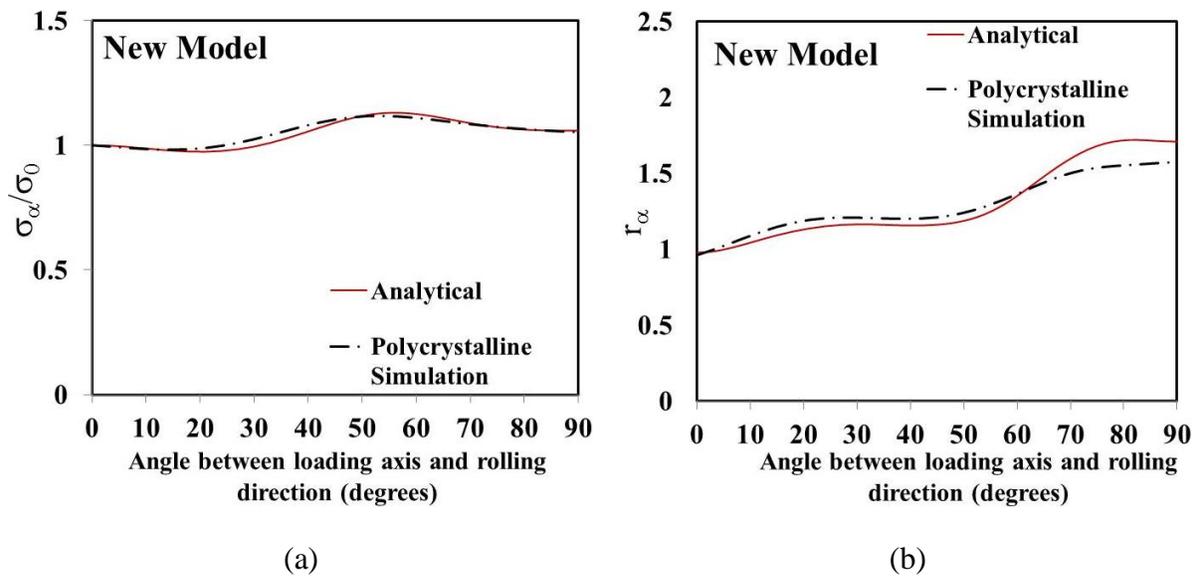

Fig. 13. Prediction of (a) yield stress ratio and (b) strain-ratio in the plane of the polycrystalline sheet predicted by the new polycrystal model for a strongly textured polycrystal with components spread around the {112}<1$\bar{1}$1>(50%) and {110}<001>(50%) orientation. The texture is shown in Figure 12.

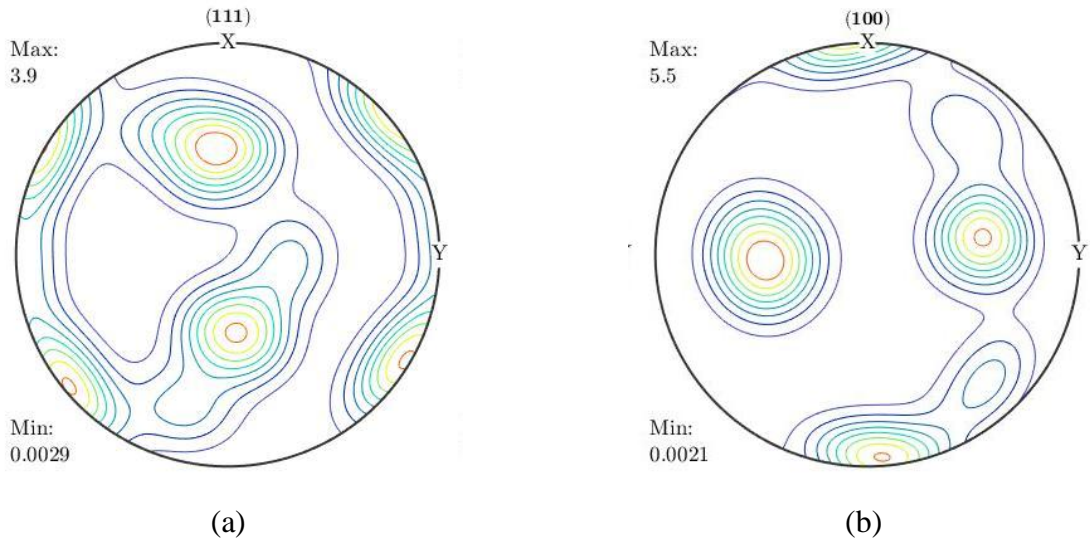

(a)                  (b)

Fig.14. Pole figures for a polycrystal with mixture of $\{\bar{2}\bar{1}1\}$<011> components (30%) and $\{110\}$<001>(70%) component (70%) (a) (111) (b) (100).

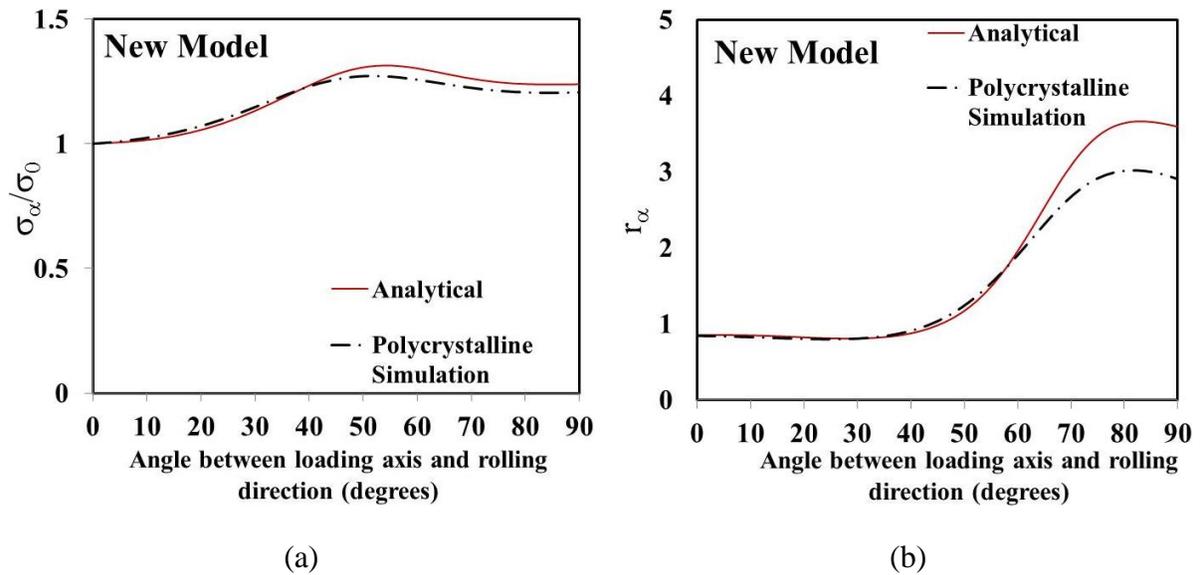

(a)                  (b)

Fig. 15. Prediction of (a) yield stress ratio and (b) strain-ratio in the plane of the polycrystalline sheet predicted by the new polycrystal model for a strongly textured polycrystal with components spread around the $\{\bar{2}\bar{1}1\}$<011>(30%) and $\{110\}$<001>(70%) orientation. The texture is shown in Figure 14.